# Low Dimensional Description of Pedestrian-Induced Oscillation of the Millennium Bridge

Mahmoud M. Abdulrehem and Edward Ott
*University of Maryland, College Park, MD 20742*

Abstract

When it opened to pedestrian traffic in the year 2000, London's Millennium Bridge exhibited an unwanted, large, side-to-side oscillation which was apparently due to a resonance between the stepping frequency of walkers and one of the bridge modes. Models for this event, and similar events on other bridges, have been proposed. The model most directly addressing the synchronization mechanism of individual walkers and the resulting global response of the bridge-pedestrian system is one developed by Eckhardt *et al*. This model treats individual walkers with a phase oscillator description and is inherently high dimensional with system dimensionality ($N+2$), where $N$ is the number of walkers. In the present work we use a method proposed by Ott and Antonsen to reduce the model of Eckhardt *et al*. to a low dimensional dynamical system, and we employ this reduced description to study the global dynamics of the bridge/pedestrian interaction. More generally, this treatment serves as an interesting example of the possibility of low dimensional macroscopic behavior in large systems of coupled oscillators.

**In order to celebrate the new Millennium, a foot bridge over the River Thames was designed and built. After an opening ceremony, an eager crowd of people streamed onto the bridge. As the number of people on the bridge increased, a relatively large lateral wobble of the bridge set in. Subsequently, the bridge was close, and a fix was designed and implemented (at considerable cost). This phenemon has also been observed in other footbridges, but is also of more general scientific interest as a clear example of emergent behavior in a large system of many coupled oscillators (where we view the individual walkers as oscillators with oscillator frequencies equal to their stepping frequencies). In this paper, we will show an inherently high dimensional microscopic description of many individual walkers self-consistently coupled to each other through the bridge response can be reduced to a low dimensional system of just a few coupled ordinary differential equations, and we use this dimension-reduced description to study the wobbling that was observed when the bridge opened.**

## I. INTRODUCTION

On opening day the Millennium Bridge, a pedestrian footbridge crossing the Thames River in London, was observed to exhibit a pronounced lateral wobbling as more and more people streamed onto the bridge [1]. This phenomenon apparently was due to a resonance between a low order bridge oscillation mode and the natural average stepping frequency of human walkers: a small initial oscillation of the bridge induces some of the walkers to synchronize the timing of their steps to that of the bridge oscillations, thus exerting a positive feedback force on the bridge that derives the bridge oscillation to higher amplitude, eventually resulting in a large steady-state oscillation. Subsequent studies by the bridge builder showed that oscillations did not develop unless the number of pedestrians on the bridge was greater than a critical value [1, 2]. After this event became widely publicized, it began to emerge that this phenomenon had also been observed on other footbridges [1, 2]. More generally, the phenomenon on these footbridges may be viewed as a particularly dramatic example of the emergence of global collective behavior in systems of many coupled heterogeneous oscillators [3] [examples of this general type of emergent behavior span biology (e.g., synchronization of pacemaker cells in the heart [4], and of neurons governing day-night rhythms in mammals [5]), chemistry (e.g. Ref. [6]), physics (e.g., Ref. [7]), etc.]

Several mathematical models have been proposed to describe the phenomenon observed on the Millennium Bridge [1, 2, 8-11]. The models of Refs. [1, 2, 8-10] are low dimensional, but do not address the issue of the mechanism by which individual walkers synchronize with the bridge, nor do they address the very important issue that different walkers have different natural stepping frequencies. Eckhardt *et al.* [10] do address these issues, and they give a detailed argument for their method of modeling the response of walkers to bridge motion. The greater degree of realism of walker modeling achieved in Ref. [10], however, results in a high dimensional model, of dimensionality (*N+2*). Here *N* is the number of pedestrians and the bridge is described by a second order ordinary differential equation (hence the + 2). Furthermore, to facilitate analysis, it proves convenient to take $N \to \infty$. The main goal of our current paper is to show how to reduce the $N \to \infty$ version of the model of Eckhardt *et al.* [10] to low dimensionality, thus producing a formulation combining the advantages of both classes of Millennium Bridge models (Refs. [1, 2, 8-10], on the one hand, and Ref. [10], on the other). Furthermore, we will utilize our reduced description to study the dynamics of the bridge/ pedestrian system.

We note that our dimension reduction makes use of the recently developed technique of Ref. [11]. This technique may be applicable to a wide variety of problems involving large systems of coupled phase oscillators, the simplest example of which is the well-known Kuramoto model [12].

The organization of this paper is as follows. In Sec. II we develop our reduced model for the case where the number of walkers on the bridge $N$ is constant in time. In Sec. III we offer analytical and numerical treatment of the reduced model developed in Sec. II. Motivated by an experimental study by the bridge builder [1, 2] in which the number of walkers on the bridge was increased in with time, we consider this situation in Sec. IV; in particular, we generalize our model of Sec. II to this case, and we numerically examine the resulting behavior.

## II. REDUCED MODEL

*a. Review of the model of Eckhardt et al.*

The model proposed in Ref. [10] describes the lateral modal bridge displacement $y(t)$ as satisfying a damped harmonic oscillator equation forced by the walkers,

$$M\ddot{y} + 2M\varepsilon\dot{y} + M\Omega^2 y = \sum_{i=1}^{N} f_i(t), \qquad (1)$$

where $M$ is the modal mass, $\varepsilon$ is the modal damping rate, $\Omega$ is the mode resonant frequency, $f_i(t)$ is the lateral modal force exerted by pedestrian $i$, and there are $N$ pedestrians on the bridge. As argued in Ref. [10], at least for low bridge oscillation amplitude, the pedestrians can be modeled as phase oscillators [12]. That is, the pedestrian force is

$$\tilde{F}_i(t) = F_i \cos[\theta_i(t)], \tag{2}$$

where $F_i$ is the peak modal force applied by walker $i$, and $\theta_i(t)$ is the phase of the walker $i$. In the absence of bridge motion, $\dot{\theta}_i = \omega_i$, where $\omega_i$ is the natural stepping frequency of walker $i$. In the presence of bridge motion, since the walkers move in the frame of the bridge, they are influenced by the inertial force due to the bridge acceleration $\ddot{y}(t)$. As a consequence, and, as argued in Ref. [10], a reasonable model of this modification is

$$\dot{\theta}_i = \omega_i - \beta_i \ddot{y} \cos[\theta_i(t)], \tag{3}$$

where $\beta_i$ is a coupling constant reflecting the strength of the response of walker $i$ to a lateral force. Data on the natural stepping frequency of humans show that the frequencies of a typical group of humans is, on average, peaked at about 1 Hz, with a spread of the order of 0.1 Hz [13]. This average stepping frequency is slightly below the resonant frequency of the lowest order lateral mode of the north span of the Millennium Bridge. The bridge damping was fairly low $\varepsilon/\Omega \ll 0.1$. The closeness of the average walker frequency, denoted $\bar{\omega}$, to the bridge resonance $\Omega$, as well as the small value of $\varepsilon/\Omega$, were used in the arguments of Ref. [10] to justify the phase oscillator walker model, Eqs. (2) and (3).

While, as we have noted, there is available data on the distribution of $\omega_i$ for a typical group of humans, we do not know of any such data for $F_i$ and $\beta_i$. Ideally, for input to this model (Eqs. (1)-(3)), we would like to have the joint probability distribution function for all these walker parameter $G(\omega, F_i, \beta)$. In the absence of such data, we will set all the $F_i$ and all the $\beta_i$ equal,

$$F_i = \bar{F} \quad , \beta_i = \bar{\beta} \tag{4}$$

but we will use a distribution function for the $\omega_i$, and we denote this distribution function $g(\omega)$. We do not view the assumption (4) as capable of making a crucial qualitative change in the phenomenon we investigate. The inclusion of a distribution of the walker frequencies is, however, essential, and, as we will later see, it plays the key role in determining the critical number of pedestrians necessary for the onset of bridge oscillation.

Concerning $\bar{F}$ and $\bar{\beta}$, referring to Ref. [10], we use as nominal values for these quantities $\bar{F} = 25N$ and $\bar{\beta} = 1.75 s/m$. We note, however, that the value estimated for $\bar{\beta}$ is only an order of magnitude estimate and that no relevant direct experimental measurements of this quantity have, to our knowledge, ever been made.

The model (1)-(3) is a dynamical system of dimensionality ($N+2$). We now assume that the number of walkers is large and approximate the ensemble of walkers using a distribution function $f(\theta, \omega, t)$ such that $f(\theta, \omega, t) d\theta d\omega$ is the fraction of walkers at time $t$ whose phases are in the interval $(\theta, \theta + d\theta)$ and whose frequencies are in the range $(\omega, \omega + d\omega)$. Formally, this description corresponds to the limit $N \to \infty$ with $N\bar{F}$ held fixed. Note that in terms of $f(\theta, \omega, t)$, the distribution function of natural frequencies $g(\omega)$ is $g(\omega) = \int_0^{2\pi} f(\theta, \omega, t) d\theta$. In this continuum limit the model (1)-(4) becomes [10]

$$M\ddot{y}(t) + 2M\varepsilon\dot{y}(t) + M\Omega^2 y(t) = N\bar{F}\,\text{Re}\big[R(t)\big], \tag{5}$$

$$R(t) = \int_{-\infty}^{\infty} d\omega \int_0^{2\pi} f(\theta,\omega,t)e^{i\theta}d\theta, \tag{6}$$

$$\frac{\partial f}{\partial t} + \omega\frac{\partial f}{\partial \theta} - \bar{\beta}\ddot{y}\frac{\partial}{\partial \theta}(f\cos\theta_i) = 0. \tag{7}$$

The quantity $R(t)$, introduced above, represents the averaged normalized walker forcing of the bridge. When stepping of the walkers is uncorrelated, the distribution function $f$ is uniform in $\theta$,

$$f = g(\omega)/2\pi, \tag{8}$$

for which (6) yields $R=0$, which is consistent with the at-rest solution of (5) and (7), $y(t)=0$. As shown by Ref. [10], this solution becomes unstable to the onset of bridge wobbling $(y(t)\neq 0)$ and synchronization of the walkers $(|R|>0)$, if $N$ exceeds a critical number of walker $N_c$, and the value found in [10] for $N_c$ was in reasonable agreement (to within inherent uncertainties) with the experimental value [1, 2] for the Millennium Bridge.

### b. Reduction of the model to low dimensionality

Following the technique of Ott and Antonsen [11], we expand $f(\theta,\omega,t)$ in a Fourier series in $\theta$,

$$f(\theta,\omega,t) = \frac{g(\omega)}{2\pi}\left\{1 + \left[\sum_{n=1}^{\infty} f_n(\omega,t)e^{in\theta} + c.c.\right]\right\}, \tag{9}$$

where c.c stands for complex conjugate. Substituting (9) into (7) and considering a restricted class of distribution functions such that

$$f_n(\omega,t) = [\alpha(\omega,t)]^n, \tag{10}$$

we obtain

$$\frac{\partial \alpha(\omega,t)}{\partial \tilde{t}} + i\left\{\omega\alpha - \frac{\bar{\beta}}{2}\ddot{\tilde{y}}[\alpha^2+1]\right\} = 0. \tag{11}$$

Now using (9) and (10) in (6), we obtain

$$R^*(t) = \int_{-\infty}^{\infty} g(\omega)\alpha(\omega,t)\,d\omega \tag{12}$$

where $R^*$ is the complex conjugate of $R$.

The special class of distribution functions given by (10) constitutes an invariant manifold $\mathfrak{M}$ in the space of distribution functions; i.e., an initial condition satisfying (10) evolves to another state satisfying (10), and it does so according to the evolution equations, (11) and (12), for $\alpha(\omega,t)$. We are interested in evolution from near the incoherent state (8) which is on $\mathfrak{M}$ (Eq. (9) and (10) with $\alpha=0$). Thus our solution on $\mathfrak{M}$ will be the relevant one, if perturbations from $\mathfrak{M}$ do not grow with time. That this is indeed the case is supported by the observation that our reduced solutions agree with those of Eqs. (1) and (3) numerically obtained in Ref. [10].

To proceed, we now adopt a convenient from the distribution function of natural walker frequencies. In particular, as in Ref. [11], we take $g(\omega)$ to be Lorentzian,

$$g(\omega) = \frac{\Delta}{\pi}\frac{1}{(\omega-\bar{\omega})^2+\Delta^2}, \tag{13}$$

where $\bar{\omega}$ is the mean walker frequency and $\Delta$ is the spread in the frequencies of the walkers.

The integral in (12) can now be done by analytically continuing $\omega$ into the complex $\omega$-plane, taking $\alpha(\omega,t)$ to be analytic and bounded in $\text{Im}(\omega) < 0$ (see Ref. [11]), and closing the integration path in the lower-half $\omega$-plane with a large semicircle of radius approaching infinity. The integral is then given by the residue of the integrand at the pole of $g(\omega)$ that is at $\omega = \bar{\omega} - i\Delta$,

$$R^*(t) = \alpha(\bar{\omega} - i\Delta, t). \tag{14}$$

Inserting $\omega = \bar{\omega} - i\Delta$ in Eq. (11), we obtain a simple first order ordinary differential equation for the evolution of complex quantity, $R(t)$,

$$\frac{dR}{d\tilde{t}} - i\left\{(\bar{\omega} + i\Delta) R^* - \frac{\bar{\beta}}{2} \ddot{\tilde{y}}\left[R^2 + 1\right]\right\} = 0. \tag{15}$$

Our exact reduction of the model of Ref. [10] to low dimension thus consists of Eqs. (5) and (15).

## III. ANALYSIS OF THE REDUCED MODEL

### a. Linear Analysis

Linearizing (5) and (15) and assuming $y \sim \exp(s\Omega t)$, $R \sim \exp(s\Omega t)$, we obtain the following equation for $s$,

$$\left[(s + \tilde{\Delta})^2 + \tilde{\omega}^2\right](s^2 + 2\tilde{\varepsilon}s + 1) = \frac{\tilde{\beta}}{2} s^2 \tilde{\omega}, \tag{16}$$

where we have introduced the normalizations,

$$\tilde{\Delta} = \Delta/\Omega, \quad \tilde{\omega} = \bar{\omega}/\Omega, \quad \tilde{\varepsilon} = \varepsilon/\Omega, \quad \tilde{\beta} = \bar{\beta}\frac{N\bar{F}}{M\Omega}. \tag{17}$$

We note that both $\tilde{\Delta}$ and $\tilde{\varepsilon}$ are small $(<0.1)$ and that $|\tilde{\omega}-1|$ is also small. Expanding (16) for

$$1 \gg |\tilde{\omega}-1| \sim \tilde{\varepsilon} \sim \tilde{\mu} \sim \sqrt{\tilde{\beta}}, \tag{18}$$

where $\tilde{\mu} = \tilde{\omega}-1$, Eq. (16) becomes

$$(\sigma + \tilde{\Delta} - i\tilde{\mu})(\sigma + \tilde{\varepsilon}) = \tilde{\beta}/8, \tag{19}$$

where $\sigma = s - i$.

We define a critical value of $\tilde{\beta}$, denoted as $\tilde{\beta}_c$, as the value of $\tilde{\beta}$ at which the real part of $\sigma$ transitions from negative to positive as $\tilde{\beta}$ increases. Once $\tilde{\beta}_c$ is known, the critical number of walkers for wobble onset is predicted to be

$$N_c = \frac{M\Omega\tilde{\beta}_c}{\overline{F}\overline{\beta}} \tag{20}$$

In the sense of giving the lowest $\tilde{\beta}_c$ (equivalently the lowest number of walkers), the worst situation is the case when the walkers' mean frequency $\overline{\omega}$ matches the bridge mode frequency $\Omega$, i.e. $\tilde{\mu} = 0$. With $\tilde{\mu} = 0$, Eq. (19) gives a critical value of $\tilde{\beta}$ of

$$\tilde{\beta}_c = 8\tilde{\Delta}\tilde{\varepsilon}, \tag{21}$$

and a solution for the instability growth rate of

$$\sigma = \frac{1}{2}\left[\sqrt{(\tilde{\Delta}+\tilde{\varepsilon})^2 - \frac{1}{2}(\tilde{\beta}-\tilde{\beta}_c)} - (\tilde{\Delta}+\tilde{\varepsilon})\right]. \tag{22}$$

If $\mu \neq 0$, then $\tilde{\beta}_c$ is given by

$$\tilde{\beta}_c/(8\tilde{\Delta}\tilde{\varepsilon}) = 1 + \frac{\tilde{\mu}^2}{(\tilde{\Delta}+\tilde{\varepsilon})^2}. \tag{23}$$

Thus the critical number of walkers increases quadratically with the difference between the mean walker frequency $\bar{\omega}$ and the bridge resonant frequency $\Omega$. Also note that inclusion of the spread in natural walker frequencies ($\tilde{\Delta} \neq 0$) is crucial, in that, without it, $\tilde{\beta}_c$ from (23) would be zero and there would thus be no threshold for instability (i.e., $N_c = 0$ from (20)).

Equations (20) and (21) with our nominal values of parameter for the Millennium Bridge and walkers ($\bar{F} = 25N$, $\bar{\beta} = 1.75 s/m$, $\Omega/2\pi = 1.03 Hz$, $M = 1.13 \times 10^5 Kg$, $\bar{\omega}/2\pi = 1.03$, $\Delta/\bar{\omega} = 0.072$) give $N_c = 73$ in the range of the observed value [1, 2].

### b. Nonlinear saturation

We now examine the saturated state for the case $\bar{\omega} = \Omega$, (i.e., $\tilde{\mu} = 0$). Using (18) we assume $y(t) = A \cos \Omega t$ and suppose that the amplitude of the component of $y(t)$ oscillating at the frequency $2\Omega$ is negligible. With this assumption, we substitute $y(t) = A \cos \Omega t$ into (15) and solve for the component of $R(t)$ that oscillators as $e^{i\Omega t}$ (we are interested in this component since it is the only one that drives $y(t)$ resonantly). Denoting this component of $R(t)$ by $i r e^{i\Omega t}$ and substituting into (15) and (5), respectively, gives

$$\Delta r = \beta \frac{A}{4}(1 - r^2), \tag{24}$$

$$2MA\varepsilon \Omega A = N \bar{F} r, \tag{25}$$

which yield the steady saturated state

$$r = \sqrt{1 - \beta_c/\beta}, \quad A = \frac{N \bar{F}}{2M\varepsilon\Omega} r. \tag{26}$$

Figure 1 shows the peak value $\tilde{y}_{max}$ of the steady state oscillation $y(t)$ normalized to $N\bar{F}/(2M\varepsilon\Omega)$ as obtained from numerical solution of Eqs. (5) and (15) plotted versus $\beta/\beta_c$ (circles). The parameters used for this plot are $\varepsilon/\Omega = 0.0075$, $\bar{\omega}/\Omega = 1$, $\Delta/\Omega = 0.072$, which are realistic representative values appropriate for the Millennium Bridge. Also plotted on Fig. 1 as a solid line is the theoretical result from (26). We see that the agreement between them is very good. The slight downward displacement of the circles from the theoretical curve is, we believe, due to our neglect of the $2\Omega$ harmonic in our derivation of (26).

## IV. TIME VARIATION OF THE NUMBER OF WALKERS

Following the discovery of the walker-induced wobble of the Millennium Bridge, the bridge builder conducted a controlled test. Using company employees, they introduced successively more walkers onto the bridge, increasing the number of walkers by the addition of groups at discrete times. The step-like curve in Fig. 2 shows the number of walkers on the bridge as a function of time. Also plotted is the sideways lateral acceleration as measured by an accelerometer attached to the bridge.

It is seen that the oscillations onset occurs rather abruptly as the number of walkers increases. When the bridge began to oscillate strongly, the people were rapidly removed from the bridge. Thus a steady state (as in Eq. (25)) was apparently not attained in this experiment.

We now wish to adapt our formulation in Sec. II to simulate this situation. Since we regard the newly introduced walkers to initially be randomly distributed in phase at the time of introduction, we will allow for a different Lorentzian distribution function for each group of walkers. Adapting our previous formulation to this situation, we have in place of (5),

$$M\ddot{y} + 2M\varepsilon\dot{y} + M\Omega^2 y = \bar{F}\sum_{j=1}^{J(t)} N_j \operatorname{Re}\left[R_j(t)\right], \tag{27}$$

where $N_j$ walkers are introduced onto the bridge at the time $t_j$, the number $J(t)$ of groups of walkers on the bridge at time $t$ is defined by

$$t_{J+1} > t \geq t_J, \tag{28}$$

and the complex quantity $R_j(t)$ characterizes the distribution of walkers in group $j$. For $t \geq t_j$, $R_j(t)$ satisfies Eq. (15) with the initial condition,

$$R_j(t_j) = 0, \tag{29}$$

corresponding to the walker phases being randomly distributed at the time when they first enter the bridge. Thus, by virtue of their different times of entry, Eq. (29) implies a different distribution of oscillator phases of each group.

Proceeding in this way, we introduce 20 groups of 10 walkers each, at a rate of 1 group per minute. The result is in rough agreement with that seem in Fig. 2.

In conclusion, we have obtained a low dimensional formulation of a previous high dimensional model for walker synchronization on the Millennium Bridge, and we have demonstrated the usefulness of our formulation for studying various aspects of the dynamics of the bridge/pedestrian system. More generally, our work serves as an interesting example of the potential for low dimensional macroscopic behavior in systems consisting of many coupled oscillators.

This work was supported by NSF and by ONR.

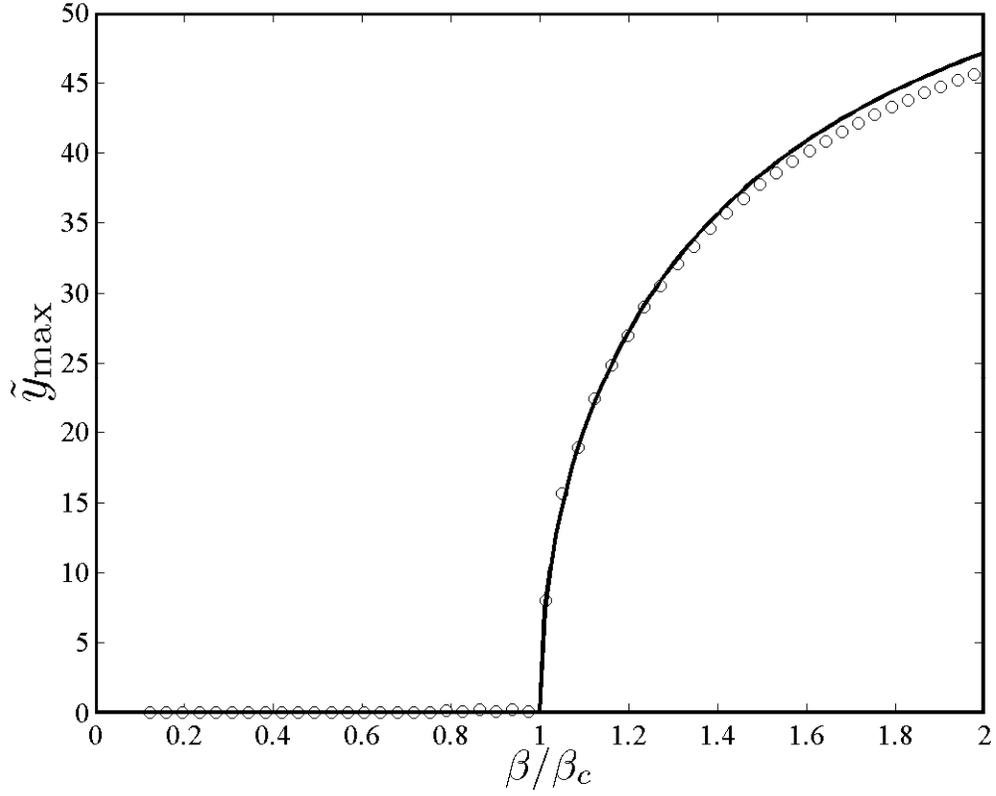

Figure 1: Plot of the normalized oscillation amplitude $\tilde{y}_{\max} = 2M\varepsilon\Omega\, y_{\max}/(N\bar{F})$ as obtained from a numerical solution of Eqs. (5) and (15) with $\varepsilon/\Omega = 0.0075$, $\bar{\omega}/\Omega = 1$, $\Delta/\Omega = 0.072$ (circles). Also plotted is the theoretical result from Eq. (26) (solid curve).

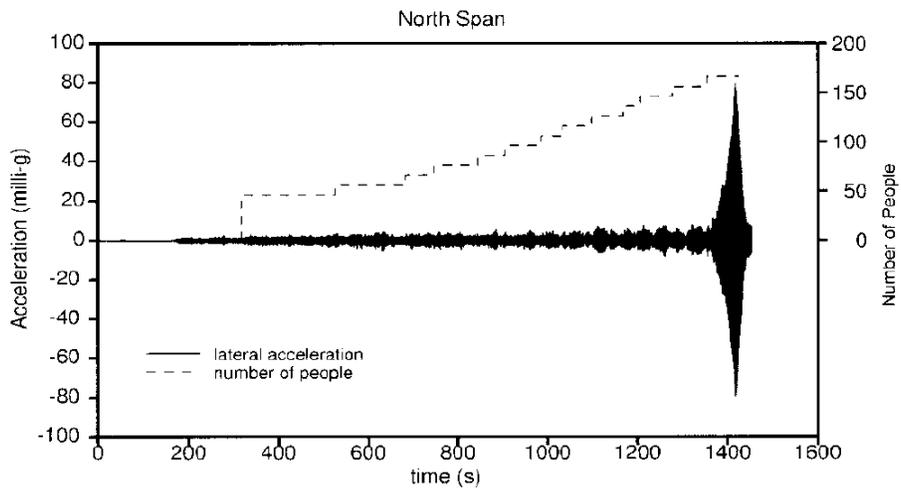

Figure 2: A time trace of lateral acceleration of the bridge deck and the number of pedestrians (taken from Arup's measurements [2]). (The numbers on the vertical axis are in units of $10^3$ of the acceleration of gravity)

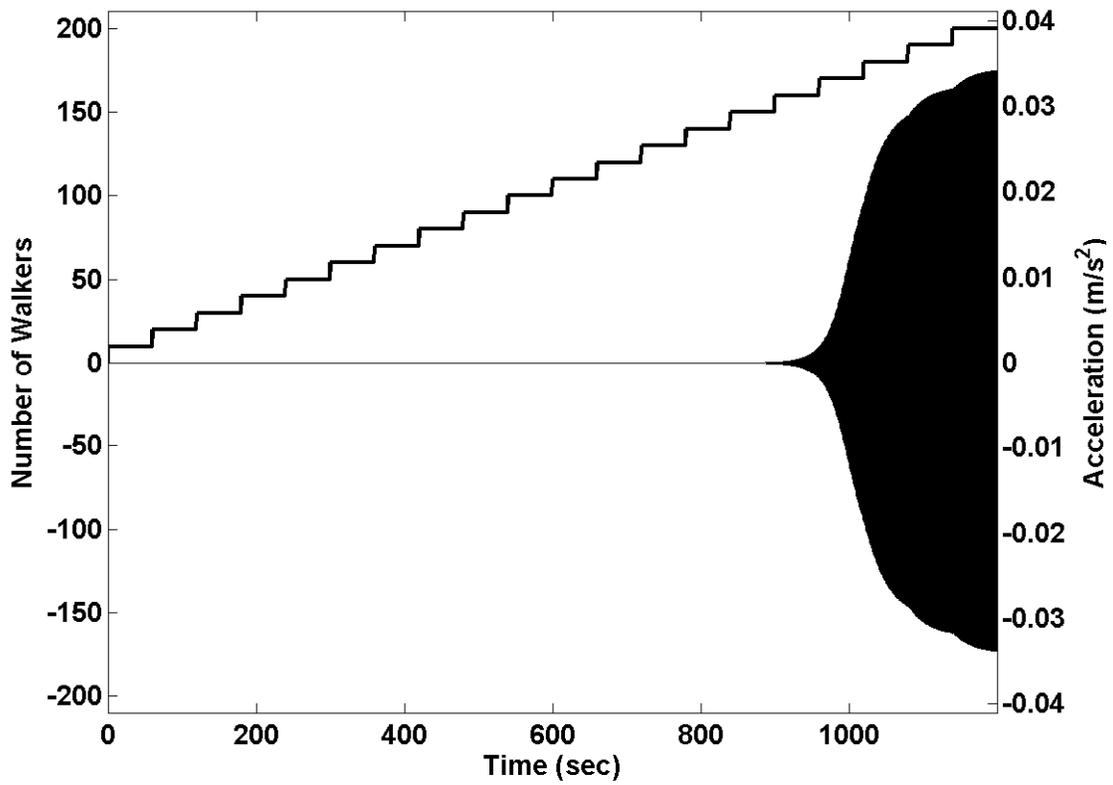

Figure 3: Effect of adding more walkers to the bridge as a function of time. The dash line is the number of walkers versus time and the solid line is the lateral acceleration versus time.